\documentclass[a4paper, 10pt]{article}
\usepackage[latin1]{inputenc}

\pdfoutput=1

\usepackage{jcappub} 

\usepackage{verbatim}
\usepackage{footnote}
\usepackage{caption}
\usepackage{subfig}
\usepackage{microtype}

\graphicspath{{figures/},
             }

\renewcommand{\d}{\mathrm{d}}
\newcommand{\p}{_{\|}}
\renewcommand{\o}{_{\perp}}

\newcommand{\lm}{^{(\ell m)}}
\newcommand{\lmp}{^{(\ell' m')}}

\newcommand{\overbar}[1]{\mkern 1.5mu\overline{\mkern-1.5mu#1\mkern-1.5mu}\mkern 1.5mu}

\title{ Light propagation in linearly perturbed $\Lambda$LTB models }

\author[a]{Sven Meyer,}
\author[a]{Matthias Bartelmann}

\affiliation[a]{Zentrum f\"ur Astronomie der Universit\"at Heidelberg, Institut f\"ur 
Theoretische Astrophysik, Philosophenweg~12, 69120 Heidelberg, Germany}

\emailAdd{ sven.meyer@uni-heidelberg.de }
\emailAdd{ bartelmann@uni-heidelberg.de }

\abstract{

We apply a generic formalism of light propagation to linearly perturbed spherically symmetric dust models including a cosmological constant. For a comoving observer on the central worldline, we derive the equation of geodesic deviation and perform a suitable spherical harmonic decomposition. This allows to map the abstract gauge-invariant perturbation variables to well-known quantities from weak gravitational lensing like convergence or cosmic shear. The resulting set of differential equations can effectively be solved by a Green's function approach leading to line-of-sight integrals sourced by the perturbation variables on the backward lightcone. The resulting spherical harmonic coefficients of the lensing observables are presented and the shear field is decomposed into its E- and B-modes. Results of this work are an essential tool to add information from linear structure formation to the analysis of spherically symmetric dust models with the purpose of testing the Copernican Principle with multiple cosmological probes.      
}


\begin{document}

\maketitle

\section{Introduction}
\label{sect:intro}

Exact cosmological solutions of general relativity (GR) have become an important tool to test the foundations of the standard cosmological model. These particular models are based on the class of spatially homogeneous and isotropic Friedmann-Lema\^{i}tre-Robertson-Walker (FLRW) models and turned out to be remarkably successful in describing multiple observational probes on a huge variety of time- and spatial scales (see for example \cite{bartelmann_dark_2010} for a review). Despite this success, its foundations need to be tested in a best possible, complete and consistent way. One possible approach focusses on the construction of more general exact solutions of GR and deriving possible observational implications. One of the simplest possible generalisations of the FLRW class is the $\Lambda$-Lema\^{i}tre-Tolman-Bondi ($\Lambda$LTB) spacetime (see \cite{lemaitre_univers_1933}, \cite{tolman_effect_1934}, and \cite{bondi_spherically_1947}) that can be foliated into spatial hypersurfaces that are spherically symmetric about one distinct central worldline. The corresponding degree of freedom of a radial density and curvature profile of the universe allows to model the possible deviations from spatial homogeneity that would break the Copernican Principle. For extensive reviews on the properties of ($\Lambda$)LTB solutions we refer to (\cite{bolejko_inhomogeneous_2011, clarkson_establishing_2012, enqvist_lemaitre_2008, marra_observational_2011}). It is important to constrain these deviations with best significance including as many as possible of the cosmological observables available. Cosmological models based on the $\Lambda$LTB solution have been constrained by multiple observational probes and so far no significant deviation from spatial homogeneity has been found (see \cite{marra_observational_2011, redlich_probing_2014}). However, up to very few exceptions based on simplifying assumptions (see \cite{moss_precision_2011, dunsby_how_2010}), a fully consistent inclusion of information from linear structure formation is still missing which excludes several important cosmological probes like cosmic shear or the integrated Sachs-Wolfe effect.\bigskip

Linear perturbation theory in radially inhomogeneous solutions are substantially more complicated than in standard FLRW models. The reduced degree of symmetry causes the dynamical evolution of gauge-invariant linear perturbations to be described by partial differential equations that contain a complicated dynamical coupling. The full evolution equations have first been derived in \cite{clarkson_perturbation_2009} while first numerical investigations were performed in \cite{february_evolution_2014, meyer_evolution_2015}. However, the structure of these gauge-invariant quantities is non-trivial as they reduce to complicated mixings of FLRW scalar-vector-tensor variables in the limit of spatial homogeneity (see \cite{clarkson_perturbation_2009} for the first detailed analysis of this issue). This means that, although the dynamics of gauge-invariant, physical perturbation variables in $\Lambda$LTB cosmologies can be modeled numerically, the results cannot be interpreted physically in a straightforward way. \bigskip   

In this context, light propagation in $\Lambda$LTB models is a promising approach to study observational effects of gauge-invariant perturbative quantities on these radially inhomogeneous backgrounds. In fact, combined influences of metric and matter perturbations on null geodesics can be mapped to corrections to the angular diameter distance that itself can be converted to observables extracted from weak gravitational lensing. This work aims at constructing the necessary expressions connecting light propagation equations to the combined effect of gauge-invariant metric and fluid perturbations. It therefore provides the foundations to include observables from linear structure formation into a most complete analysis of $\Lambda$LTB models. \bigskip
  
The paper is structured as follows: Sect. (\ref{sect:generic}) outlines a generic and well-known relativistic approach to light propagation starting with thin bundles of null geodesics. A short summary on the geodesic deviation equation in $\Lambda$LTB models is provided in Sect. (\ref{sect:background}). In the following Sect. (\ref{sect:perturbed}), we derive the full equation system for geodesic deviation in linearly perturbed $\Lambda$LTB models which is decomposed into spherical harmonics functions. In Sect. (\ref{sect:greens}) we address a possible solution based on a Green's function approach yielding line-of-sight integral expressions for the lensing observables. The resulting cosmic shear field will then be split into the E- and B-modes in Sect. (\ref{sect:EB-modes}).

\section{Light propagation in general relativity}
\label{sect:generic}

The following section provides a short summary on relativistic light propagation as well as the conventions and notation applied in this work. It is mainly based on the approaches presented in \cite{bartelmann_topical_2010, clarkson_misinterpreting_2012}.  We consider an infinitesimal bundle of null geodesics (see \cite{perlick_gravitational_2010} for an exact definition) that is propagating in an arbitrary spacetime and converges at an observer freely falling with four-velocity $u_\mathrm{obs}$. One particular geodesic of the bundle can be singled out as a so-called \textit{fiducial ray} and parametrised by an \textit{affine parameter} $\lambda$. Given the observer's local coordinates $x^\mu$, we define the ray's wave vector as

\begin{equation}
 k^\mu = \frac{\d x^\mu}{\d \lambda}
 \label{lltb:generic:1}
\end{equation}

and choose $\lambda$ such that a unit projection of $k$ on $u_\mathrm{obs}$ is obtained. Effectively, this corresponds to a normalisation of the wave vector by the observed frequency of the light ray. Starting from 

\begin{equation}
 \langle k, u_\mathrm{obs} \rangle = -\omega_\mathrm{obs}\,,
  \label{lltb:generic:2}
\end{equation}

we transform $k^\mu \longrightarrow \tilde{k}^\mu = -k^\mu/\omega_\mathrm{obs} \equiv k^\mu$ such that

\begin{equation}
 \langle k, u_\mathrm{obs} \rangle = 1\,.  
  \label{lltb:generic:3}
\end{equation}

Given this affine parametrisation, $\lambda$ corresponds to the Euclidian distance in the local neighborhood of the freely falling observer $\d \lambda = \d r$. In addition, the redshift of a fictitious source with respect to the observer can be defined as  
 
 \begin{equation}
    \langle k, u_\mathrm{s} \rangle = 1 + z\,, 
    \label{lltb:generic:4}
 \end{equation}

 where $u_\mathrm{s}$ denotes the source's four-velocity. The redshift is normalised to zero for a comoving source placed at the observer's position. \bigskip

 We now consider the spacelike plane perpendicular to $k$ and $u_\mathrm{obs}$ which defines a screen in the rest frame of the observer. An orthonormal basis of this screen is generally given by the two vectors $n^\mu_{\ a}$ $(a=1,2)$ which are commonly referred to as \textit{Sachs basis}. By construction, the Sachs basis vectors then fulfill the following identities\footnote{We will denote the observer's four velocity as $u^\mu$ in the following and drop the subscript.}:
 
 \begin{align}
  \label{lltb:generic:5}
  k_\mu {n^\mu}_a  &= 0\,, \\
  \label{lltb:generic:6}
  u_\mu {n^\mu}_a &= 0\,, \\
  \label{lltb:generic:7}
  n_{\mu,a \vphantom{b}} {n^\mu}_b &= \delta_{ab}\,. 
 \end{align}
 
 Having set-up the Sachs basis at $\lambda=0$, the basis vectors at arbitrary affine parameters can be obtained by parallel transport ($\nabla_k n_a = 0$) of the initial basis along the fiducial ray. Given the Riemannian connection, Eqs. (\ref{lltb:generic:5}) - (\ref{lltb:generic:7}) are not affected by this procedure.\bigskip    
  
\begin{figure}
  \centering
  \includegraphics[width=0.5\hsize]{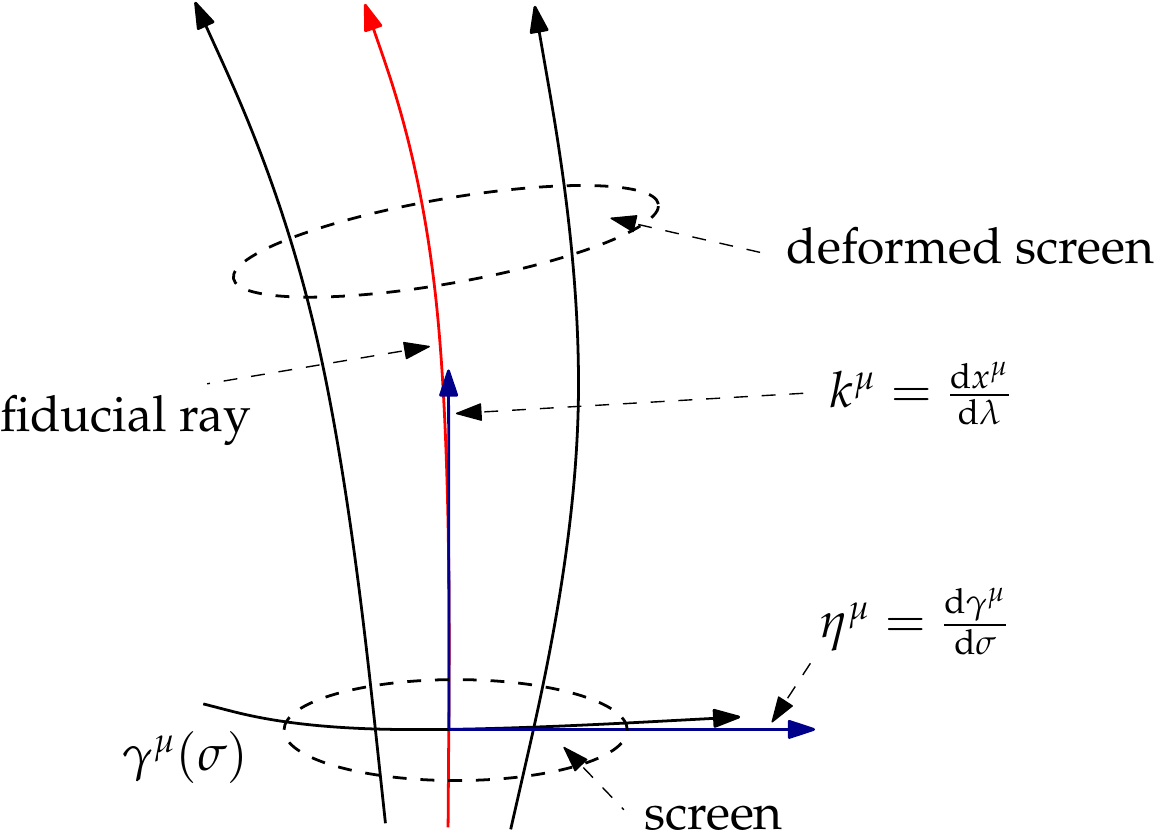}
  \caption[Construction of a screen space for a bundle of null geodesics]{Construction of a screen space for a bundle of null geodesics: The evolution of the geodesic bundle can be mapped to the corresponding deformation of the screen when parallel-transported along the fiducial ray. }
  \label{lltb:generic:fig:1}
 \end{figure}

 A general vector in the screen space can be constructed by defining a second affine parameter $\sigma$ and a corresponding spacelike curve $\gamma(\sigma)$ that connects the fiducial ray with neighboring geodesics (see Fig. (\ref{lltb:generic:fig:1})). By assumption, $\gamma(\sigma)$ is entirely contained in the screen space such that the tangent vector 
 
 \begin{equation}
 \eta^\mu = \left. \frac{\d \gamma^\mu}{\d \sigma} \right|_{\sigma=0}
  \label{lltb:generic:8}
 \end{equation}

 can be expanded into the Sachs basis vectors
 
 \begin{equation}
  \eta^\mu = \eta_1 {n^\mu}_1 + \eta_2 {n^\mu}_2\,.
  \label{lltb:generic:9}
 \end{equation}

 For a proper choice of the affine parameter $\sigma$, $\eta^\mu$ measures the physical size and shape of the bundle when parallel-transported along the fiducial ray. The evolution of $\eta^\mu$ is given by the equation of geodesic deviation 
 
 \begin{equation}
  k^\alpha k^\beta \nabla_\alpha \nabla_\beta  \eta^\mu = R^\mu_{\ \nu \alpha \beta} k^\nu k^\alpha \eta^\beta\,,
  \label{lltb:generic:10}
 \end{equation}

 containing the generic Riemann tensor of the spacetime.\bigskip
 
 Inserting Eq. (\ref{lltb:generic:9}) into Eq. (\ref{lltb:generic:10}), we obtain
 
 \begin{equation}
 \frac{\d^2 \eta_a}{\d \lambda^2} = R_{\mu\nu\alpha\beta} {n^\mu}_a k^\nu k^\alpha {n^\beta}_b \eta_b = \mathcal{T}_{ab} \eta_b\,,
 \label{lltb:generic:11} 
 \end{equation}

 where summation over $b$ is implied. The object $\mathcal{T}_{ab}$ is the so-called \textit{optical tidal matrix} as it connects the evolution of the geodesic bundle with the curvature of spacetime. It separates into two distinct contributions
 
 \begin{equation}
 \mathcal{T}_{ab} = -\frac{1}{2} R_{\alpha \beta} k^\alpha k^\beta \delta_{ab} + C_{\mu \nu \alpha \beta} {n^\mu}_a k^\nu k^\alpha {n^\beta}_b\,,
 \label{lltb:generic:12} 
 \end{equation}

 which define the so-called \textit{Ricci-} and \textit{Weyl focussing} terms. The Ricci focussing originates from matter inside the bundle that causes $\eta^\mu$ to increase or decrease isotropically. On the other hand, the Weyl focussing is generated by matter located outside the bundle giving rise to shear effects on the screen. The role of the two different contributions will be discussed below in more detail.\bigskip
 
 Since Eq. (\ref{lltb:generic:11}) is a second order ordinary differential equation in the affine parameter $\lambda$, any solution is constrained by two initial conditions given by the initial value and the initial first derivative of $\eta^\mu$. As assumed a priori, the bundle converges at the freely falling observer placed at $\lambda=0$ which fixes $\eta^\mu(\lambda=0)$ to zero. The final solution can therefore only depend on the initial rate $\left. \d\eta^\mu/\d\lambda \right|_{\lambda=0}$. In case of a linear differential equation, the solution this yields the mapping
 
 \begin{equation}
  \eta_a(\lambda) = \left. D_{ab}(\lambda) \frac{\d \eta_b}{\d \lambda}\right|_{\lambda=0}\,,
  \label{lltb:generic:13}
 \end{equation}
 
 with the \textit{Jacobi map} $D_{ab}$ that contains all information on the evolution of the geodesic bundle with respect to $\lambda$. Hence, the full initial value problem can be formulated in terms of the Jacobi map which yields Jacobi matrix equation:
 
 \begin{equation}
   \begin{split}
     \frac{\d^2 D_{ab}}{\d \lambda^2} \quad  &= \mathcal{T}_{ac\vphantom{b}} D_{cb}\,, \\  
     \left. D_{ab}\right|_{\lambda=0} &= 0\,, \\
     \left. \frac{ \d D_{ab}}{ \d \lambda }\right|_{\lambda=0} &= \delta_{ab}\,, 
   \end{split}
  \label{lltb:generic:14}
 \end{equation}
 
 which is independent of the initial rate of $\eta_a$.  We have chosen the affine parameter $\lambda$ to coincide with the local Euclidian distance in the observer's rest frame. Thus, the initial rate can locally be interpreted as the opening angle 
 
 \begin{equation}
  \theta_a = \left. \frac{\d \eta_a}{\d \lambda} \right|_{\lambda=0}
  \label{lltb:generic:15}
 \end{equation}

 in this particular frame. Integrating Eq. (\ref{lltb:generic:14}) from the observer to a fiducial source located at a position corresponding to the affine parameter $\lambda_s$ leads to 
 
 \begin{equation}
  \eta_a(\lambda_s) = D_{ab} (\lambda_s) \left. \frac{\d \eta_b}{\d \lambda} \right|_{\lambda=0} = D_{ab} (\lambda_s) \theta_b\,.
  \label{lltb:generic:16}
 \end{equation}

 This means that the Jacobi map relates cross-sectional diameters of the bundle at the source position to angular diameters at the observer which defines an angular diameter distance. Precisely, this definition only holds for infinitesimal bundles with circular cross section. In case of general elliptical cross sections, $D_{ab}$ can be diagonalised yielding two extremal angular diameter distances $D_+(\lambda_s)$ and $D_-(\lambda_s)$. In fact, a circular image of angular size $\theta$ seen by an observer has an elliptical cross-section with principal axes $|D_\pm(\lambda_s)| \cdot \theta$ at the source position (see \cite{perlick_gravitational_2010} for details). Therefore, the angular diameter distance shall be replaced by the so-called area distance that relates the cross-sectional area of the bundle at the source position to the solid angle seen by the observer. Involving the geometric interpretation of the determinant, the area distance can be defined as (see \cite{clarkson_misinterpreting_2012, perlick_gravitational_2010})
 
 \begin{equation}
  D_A(\lambda_s) = \sqrt{(\det{D_{ab}})(\lambda_s)} = \sqrt{ D_+(\lambda_s)D_-(\lambda_s)}\,.
  \label{lltb:generic:17}
 \end{equation}

Due to its general applicability, this definition will be considered as angular diameter distance in the following. $D_A$ is an important physical quantity as it can directly be inferred from observations. Once a physical length scale of a particular source is known, the opening angle can be measured and $D_A$ readily estimated. On the other hand, $D_A$ is related to the Jacobi map which is itself a solution to the Jacobi matrix equation. It is therefore sensitive to the spacetime geometry due to the Weyl and Ricci focussing terms in the optical tidal matrix. Effects of gauge-invariant perturbations of the background spacetime can therefore be mapped to physically meaningful observables. This is a most welcome property in case of more abstract gauge-invariants such as those appearing in gauge-invariant $\Lambda$LTB perturbation theory. \bigskip

The Jacobi map can be related to the Jacobian matrix $A_{ab}$ of the lens mapping (see \cite{bartelmann_topical_2010}) which is also denoted as \textit{lensing amplification matrix}. We recover again Eq. (\ref{lltb:generic:15}) since it defines the angle under which a source is seen at the observer's position. The angular position $\beta_a$ of the source without focussing effects is given by

\begin{equation}
  \beta_a = \frac{\eta_a(\lambda_s)}{\overbar{D}_A(\lambda_s)}\,,
  \label{lltb:generic:18}
 \end{equation}
 
 where $\overbar{D}_A(\lambda_s)$ is the area angular diameter distance of a background spacetime in which focussing effects due to perturbations are studied. When combining Eqs. (\ref{lltb:generic:15}) and (\ref{lltb:generic:17}), we obtain the lens map that relates the angular position of the source to the observed angular position due to focussing effects:
 
 \begin{equation}
  \beta_a = \frac{\eta_a(\lambda_s)}{\overbar{D}_A(\lambda_s)} = \frac{D_{ab}(\lambda_s)}{\overbar{D}_A(\lambda_s)} \theta_b = A_{ab}(\lambda_s) \theta_b. 
  \label{lltb:generic:19}
 \end{equation}

 Hence, the lensing amplification matrix is generally expressed as
 
 \begin{equation}
  A_{ab}(\lambda) = \frac{D_{ab}(\lambda)}{\overbar{D}_A(\lambda)}\,,
  \label{lltb:generic:20}
 \end{equation}

 which can conveniently be decomposed into a trace and trace-free part 
 
 \begin{equation}
  (A_{ab})= \left(\begin{array}{cc}
                   1-\kappa & 0 \\
                   0 & 1-\kappa
                  \end{array}
 \right) - 
 \left(\begin{array}{cc}
                   \gamma_1 & \gamma_2 \\
                   \gamma_2 & -\gamma_1
                  \end{array}
 \right)\,.
  \label{lltb:generic:21}
 \end{equation}

\section{Geodesic deviation in \texorpdfstring{$\Lambda$}\ LTB cosmologies}
\label{sect:background}
  
The concepts introduced in the previous section can now readily be applied to $\Lambda$LTB models. The $\Lambda$LTB solution is a dust solution of Einstein's field equations with hypersurfaces that are spherically symmetric about one central worldline. The line element in comoving synchronous coordinates (see \cite{straumann_general_2013}) reads

\begin{equation}
 \d s^2 = -\d t^2 + \frac{a\p^2(t,r)}{1-\kappa(r) r^2} \d r^2 + r^2 a\o^2(t,r) \d \Omega^2\,, 
 \label{lltb:background:1}
 \end{equation}
 
 with an energy momentum tensor $T_{\mu\nu} = \rho(t,r) u_\mu u_\nu$. Inward radial null geodesics for a central observer are constrained by the following equation system
 
 \begin{align}
 \label{lltb:background:2}
 \frac{\d t(r)}{\d r} &= - \frac{a\p(t(r),r)}{\sqrt{1-\kappa(r) r^2}}\,,\\
 \label{lltb:background:3}
 \frac{1}{ 1 + z(r)} \frac{\d z(r)}{\d r} &=  \frac{\dot{a}\p(t(r),r)}{\sqrt{1-\kappa(r) r^2}}\,,
 \end{align}
 
 Throughout this work, we will restrict ourselves to observers located at the center of a $\Lambda$LTB patch, because this yields a considerable simplification of the expressions derived in the next section. However, conceptually there is no restriction of the observer's position to the center. Off-center observers in LTB void models at the background level have been considered in previous works (see \cite{brouzakis_effect_2007}, \cite{fanizza_lensing_2015} as well as \cite{cusin_are_2017}). In this context, geodesic lightcone coordinates (see \cite{fleury_geodesic-light-cone_2016}) have proven to be very effective, but this approach will not be followed in this work \footnote{Although an extension to off-center observers is desirable for future considerations, it turns out that, on the one hand, CMB observations constrain the observer's position to be very close ($\sim$ few Mpc) to the center (see \cite{alnes_cmb_2006}) in case of LTB solutions and, on the other hand, deviations from spatial homogeneity in $\Lambda$LTB models are very small.}. \bigskip  
  
 Using Eqs. (\ref{lltb:background:2}), (\ref{lltb:background:3}), and (\ref{lltb:generic:1}), differential relations between the affine parameter $\lambda$ and coordinate time, radius, and redshift can be derived

  \begin{align}
 \label{lltb:background:4}
 \frac{\d t(\lambda)}{\d \lambda} &= -(1+z(\lambda))\,,\\
 \label{lltb:background:5}
 \frac{\d r(\lambda)}{\d \lambda} &= (1+z(\lambda)) \, \frac{\sqrt{1-\kappa(r(\lambda))r^2(\lambda)}}{a\p(t(\lambda), r(\lambda))}\,, \\
 \label{lltb:background:6}
 \frac{\d z(\lambda)}{\d \lambda} &= (1+z(\lambda))^2 H\p(t(\lambda), r(\lambda))\,,
 \end{align} 
 
which form a coupled system of ordinary differential equations constraining the shape of the background $\Lambda$LTB past null cone. \bigskip
   
 As the $\Lambda$LTB solution is spherically symmetric about the central worldline, the Weyl focussing term in the optical tidal matrix vanishes and the field equations constrain the Ricci focussing term to be 
 
 \begin{equation}
  \begin{split}
  \mathcal{T}_{ab} &= -\frac{1}{2} R_{\alpha \beta} k^\alpha k^\beta \delta_{ab} = -\frac{1}{2} T_{\alpha \beta} k^\alpha k^\beta \delta_{ab} \\
                   &= -4\pi G \rho(t,r) \, (1+z)^2 \delta_{ab}\,.
  \end{split}
  \label{lltb:background:7}
 \end{equation}
 
 The geodesic deviation equation then reads
 
 \begin{equation}
  \frac{\d^2 D_{ab}(\lambda)}{\d \lambda^2} = -4\pi G \rho(t(\lambda),r(\lambda)) \, (1+z(\lambda))^2 D_{ab}(\lambda)\,, 
  \label{lltb:background:8}
 \end{equation}

 which has to be solved in combination with Eqs. (\ref{lltb:background:4})-(\ref{lltb:background:6}) using the initial conditions
 
 \begin{align*}
 \begin{split}
   \left. D_{ab}\right|_{\lambda=0} &= 0\,, \\ 
   \left. \frac{\d D_{ab}}{\d \lambda} \right|_{\lambda=0} &= \delta_{ab}\,,   
 \end{split}\\
   \left. t\right|_{\lambda=0} &= t_\mathrm{age}\,, \\
   \left. r \right|_{\lambda=0} &= 0\,, \\
   \left. z \right|_{\lambda=0} &= 0\,.
 \end{align*}
  
  In general, this system has to be evolved numerically, but there exists an analytic solution to Eq. (\ref{lltb:background:8}) without knowledge of the exact shape of the backward lightcone. By taking Eqs. (\ref{lltb:background:4})-(\ref{lltb:background:6}) as differential relations and involving the field equations at the background level (see also Appendix (\ref{app:solution})), it can be shown that 
  
  \begin{equation}
   D_{ab}(\lambda) = r(\lambda) a\o(t(\lambda), r(\lambda)) \delta_{ab} 
  \label{lltb:background:9}
  \end{equation}
  
  solves Eq. (\ref{lltb:background:8}). This result which has also previously been found in (\cite{dunsby_how_2010}) is physically meaningful as it describes the areal radius of the line element (Eq. (\ref{lltb:background:8})) fixing the angular diameter distance for a central observer. Eq. (\ref{lltb:background:9}) turns out to be very useful in the following sections.

\section{Geodesic deviation in perturbed \texorpdfstring{$\Lambda$}\ LTB cosmologies}
\label{sect:perturbed}

As proposed in (\cite{gerlach_gauge-invariant_1979}), a 2+2 split of the full spacetime $\mathcal{M}^4 = \mathcal{M}^2 \times \mathcal{S}^2$ leads to gauge-invariant linear perturbations that can be expressed in terms of scalar, vector and tensor spherical harmonics. Those naturally split into an even, polar and an odd, axial branch by considering their curl-free and divergence-free parts on $\mathcal{S}^2$, respectively. In the polar branch, there are four degrees of freedom $\chi\lm$, $\varphi\lm$, $\varsigma\lm$, and $\eta\lm$ entering the linearly perturbed metric as well as three expressions $\Delta\lm$, $w\lm$, and $v\lm$ fixing the energy-momentum tensor (see \cite{clarkson_perturbation_2009}). In Regge-Wheeler (RW) gauge (see \cite{regge_stability_1957}), the perturbed metric and energy-momentum tensor read:

 \begin{align}
  \label{lltb:perturbation:1}
  ds^2 &= -\left[ 1 + (2\eta\lm - \chi\lm - \varphi\lm) Y\lm \right]\d t^2 - \frac{2 a\p \varsigma\lm  Y\lm }{\sqrt{1 - \kappa r^2}} \d t \d r  \\
       & \ \ \ \ + \frac{a\p^2}{1 - \kappa r^2} \left[ 1 + (\chi\lm + \varphi\lm) Y\lm \right] \d r^2  + r^2 a\o^2 \left[1 + \varphi\lm  Y\lm \right] \d \Omega^2 \,, \nonumber \\ \nonumber \\ 
  \label{lltb:perturbation:2}
  \rho &= \rho^\mathrm{LTB} \left( 1 + \Delta\lm  Y\lm \right)\,, \\
  \label{lltb:perturbation:3}
  u_\mu &= \left[ u_A + \left( w\lm n_A + \frac{1}{2} k_{AB} u^B  \right) Y\lm , v\lm  Y_b\lm  \right] \,, 
 \end{align}
 
 with sums over $(\ell, m)$ implied and $ Y_b^{(\ell m)} = \nabla_b  Y^{(\ell m)}$\footnote{There are three types of indices appearing in the 2+2 split of the spacetime. By convention of \cite{clarkson_perturbation_2009}, we use Greek indices for the full spacetime coordinates, capital Roman letters for the $(t,r)$-submanifold $\mathcal{M}^2$ and small Roman letters for the angular parts on $\mathcal{S}^2$.}. The unit vectors in time and radial directions are given by $u_A = (-1,0)$ and $n_A = (0, a\p/\sqrt{1-\kappa r^2})$. $k_{AB}$ corresponds to the metric perturbation in the $(t,r)$-submanifold. \bigskip
 
 The gauge-invariant perturbations of the axial branch consist of a vector field $k_A$ and scalar $\overbar{v}\lm$:
 
 \begin{equation}
 \d s^2 = -\d t^2 + \frac{a\p(t,r)^2}{1-\kappa(r)r^2} \d r^2 + r^2 a\o^2(t,r) \d \Omega^2 + 2 k_A \d x^A \bar{Y}\lm_b \d x^b\,,
 \label{lltb:perturbation:6}
 \end{equation}
 
 and 
 
 \begin{equation}
  u_\mu = \left(u_A, \bar{v} \bar{Y}\lm_a\right)\,.
  \label{lltb:perturbation:7}
 \end{equation}
 
 We arrive at six degrees of freedom in total for the metric and four degrees of freedom for the energy-momentum tensor\footnote{The latter is caused by the absence of anisotropic stress to first order such that only four of the expected six independent quantities remain in the energy-momentum tensor.}. Einstein's field equations constrain the dynamical evolution of these equations for the polar and axial branch. Whereas both branches are dynamically decoupled, this does not hold for the gauge-invariant quantities in each branch due to the reduced degree of symmetry of the $\Lambda$LTB solution with respect to FLRW models. This also leads to a complicated structure of these quantities in the limit of spatial homogeneity as they mix FLRW scalar-vector-tensor\footnote{expressed in conformal Newtonian gauge} degrees of freedom. For details on the evolution equations and construction of gauge-invariant quantities we refer to (\cite{clarkson_perturbation_2009}). First numerical investigations on the evolution of gauge-invariant quantities can be found in (\cite{february_evolution_2014}) and (\cite{meyer_evolution_2015}). \bigskip
 
 Generically perturbed $\Lambda$LTB spacetimes do not obey any symmetries seen by observers moving on the $\Lambda$LTB central worldline. Strictly speaking, even this special position in spacetime cannot be precisely singled out anymore. However, assuming that deviations from the spherically symmetric $\Lambda$LTB solution are small, the following approximations can be made:
 
 \begin{itemize}
  \item The observer's worldline is approximated by a geodesic in the background LTB spacetime. Hence, the observer's rest frame and the corresponding central worldline can be described by the $\Lambda$LTB background expressions only.
  \item Born's approximation can be applied where influences of perturbations on the propagation of null geodesics are integrated along the unperturbed lightpath. Since metric potentials are assumed to be small, this approximation is typically very accurate (see \cite{bernardeau_full-sky_2010, schafer_validity_2012}). 
 \end{itemize}

 Referring to these approximations, perturbations of the wave vector $k^\mu$ and the Sachs basis $n^\mu_{\ a}$ as well as deviations in the affine parameter $\lambda$, redshift, and lightcone coordinates from their background values  are not considered. This allows to adopt Eqs. (\ref{lltb:background:4}) - (\ref{lltb:background:6}) right away from the background model and consider only perturbations in the optical tidal matrix\footnote{We decided to keep the full metric for contractions performed in Eq. (\ref{lltb:generic:12}) and for the null condition in $k$ as this leads to Weyl focussing terms that are trace-free objects in terms of the Sachs basis $n_a$.}. \bigskip
 
 Since we deal with a spherically symmetric solution around a central observer, it is convenient to adapt the Sachs basis of the screen to a spherical basis by demanding
 
 \begin{align}
  \label{lltb:perturbation:4}
  n_{\mu,a \vphantom{b}} {n^\mu}_b &= \gamma_{ab}\,. 
 \end{align}
 
 with $(\gamma_{ab})$ denoting the metric on $\mathcal{S}^2$.\bigskip
 
 The optical tidal matrix in this screen basis then reads
 
 \begin{equation}
 \mathcal{T}_{ab} = -\frac{1}{2} R_{\alpha \beta} k^\alpha k^\beta \gamma_{ab} + C_{\mu \nu \alpha \beta} {n^\mu}_a k^\nu k^\alpha {n^\beta}_b\,.
 \label{lltb:perturbation:5} 
 \end{equation} 

 The Jacobi matrix can be split into a background contribution $D^{(0)}_{ab}$ and a linear correction $D^{(1)}_{ab}$. A similar split can be performed for the optical tidal matrix $\mathcal{T}_{ab}$. As a result, Eq. (\ref{lltb:generic:14}) becomes a coupled differential equation system
 
  \begin{align}
   \label{lltb:perturbation:8}
  \frac{\d^2 D_{ab}^{(0)}(\lambda)}{\d \lambda^2} &= \mathcal{T}_{ac \vphantom{b}}^{(0)}(\lambda) \gamma^{cd} D_{db}^{(0)}(\lambda)\,, \\
   \label{lltb:perturbation:9}
  \frac{\d^2 D_{ab}^{(1)}(\lambda, \theta, \phi)}{\d \lambda^2} &= \mathcal{T}_{ac\vphantom{b}}^{(1)}(\lambda, \theta, \phi) \gamma^{cd} D_{db}^{(0)}(\lambda) + \mathcal{T}_{ac\vphantom{b}}^{(0)}(\lambda) \gamma^{cd} D_{db}^{(1)}(\lambda, \theta, \phi)\,. 
 \end{align}
 
 By construction, contractions over angular coordinates $(a,b)$ are now performed using the metric on $\mathcal{S}^2$. By construction, Eq. (\ref{lltb:perturbation:8}) is identical to Eq. (\ref{lltb:background:8}) and its solution given by (\ref{lltb:background:9}) now reads
 
   \begin{equation}
   D^{(0)}_{ab}(\lambda) = r(\lambda) a\o(t(\lambda), r(\lambda)) \gamma_{ab}\,. 
  \label{lltb:perturbation:10}
  \end{equation}
    
  For the polar branch, we find the following expressions for the Ricci- and Weyl focussing terms sourced by polar gauge-invariant linear perturbations:
  
   \begin{align}
  \label{lltb:perturbation:11}
  \begin{split}
  R^{(1)}_{\alpha\beta} k^\alpha k^\beta &= \alpha \left( 1+z\right)^2 \sum_{(\ell m)}  \left( 2w\lm + \Delta\lm + 2\eta\lm \right.  \\
                                              & \left. \quad - \chi\lm - \varphi\lm - \varsigma\lm \right) Y\lm   \,,   
  \end{split}\\
  \label{lltb:perturbation:12}
   C^{(1)}_{\alpha\beta\gamma\delta} n_{\ a}^\alpha k^\beta k^\gamma n_{\ b}^\delta &= -\frac{\left(1+z\right)^2}{r^2 a\o^2} \sum_{(\ell m)}{\left( \eta\lm - \chi\lm - \varphi\lm - \varsigma\lm  \right) Y\lm_{ab} } 
  \end{align}
  
  with $\alpha= 8\pi G\rho(t,r)$ and the polar tensor spherical harmonic function defined as
  
  \begin{equation}
   Y\lm_{ab} = \left( \nabla_a \nabla_b + \frac{\ell(\ell+1)}{2}\gamma_{ab} \right) \ Y\lm\,.
   \label{lltb:perturbation:13}
  \end{equation}
  
   In case of the axial branch, there is no contribution to Ricci focussing since the axial fluid perturbation $\overbar{v}\lm$ only contributes to the angular components of the energy-momentum tensor and therefore does not affect  radial null geodesics of an observer comoving with the central worldline:
  
   \begin{align}
   \label{lltb:perturbation:15}
    R^{(1)}_{\alpha\beta} k^\alpha k^\beta &= 0 \,, \\   
   \label{lltb:perturbation:16}
    C^{(1)}_{\alpha\beta\gamma\delta} n_{\ a}^\alpha k^\beta k^\gamma n_{\ b}^\delta &= -\frac{\left(1+z\right)^2}{r^2 a\o^2} v^{AB} \sum_{(\ell m)}{ \nabla_A^{\vphantom{(\ell m)}} k_B\lm \overbar{Y}\lm_{ab} }\,. 
   \end{align}
  
  $Y\lm_{ab}$ denotes the axial tensor spherical harmonic function given by\footnote{Note that we decided to define the axial tensor spherical harmonics including an additional factor of $1/2$ with respect to the definition used in   \cite{clarkson_perturbation_2009}). This simplifies relations to spin-weighted spherical harmonics considered below.}
     
   \begin{equation}
   \overbar{Y}\lm_{ab} = \frac{1}{2} \left( \nabla_a^{\vphantom{(\ell m)}} \overbar{Y}_b\lm + \nabla_b^{\vphantom{(\ell m)}} \overbar{Y}_a\lm \right) \,,
   \label{lltb:perturbation:17}
  \end{equation}
  
  with $\overbar{Y}\lm_{a} = \epsilon_a^{\ b} \nabla_b Y\lm$. The tensor field $v^{AB}$ on $\mathcal{M}^2$ can be expressed in terms of the unit vectors $u_A$ and $n_A$ defined above:
  
  \begin{equation}
  v_{AB} = u_Au_B + n_A n_B - (u_An_B+n_Au_B)\,.
   \label{lltb:perturbation:18}
  \end{equation}

  Inserting Eqs. (\ref{lltb:perturbation:10})-(\ref{lltb:perturbation:12}) and (\ref{lltb:perturbation:15})-(\ref{lltb:perturbation:16}) into Eq. (\ref{lltb:perturbation:9}), we find the full expression of the first order correction to the Jacobi map $D^{(1)}_{ab}$:
  
  \begin{equation}
  \label{lltb:perturbation:19}
   \begin{split}
   \frac{\d^2 D_{ab}^{(1)}}{\d \lambda^2} &= - \left( 1+z\right)^2 \frac{\alpha}{2}\ D^{(1)}_{ab} \\
                                          & - \left( 1+z\right)^2 ra\o \frac{\alpha}{2} \sum_{(\ell m)} \left( 2w\lm + \Delta\lm + 2\eta\lm \right.  \\
                                          & \left. \quad - \chi\lm - \varphi\lm - \varsigma\lm \right) Y\lm \\
                                          & - \frac{\left(1+z\right)^2}{r a\o} \sum_{(\ell m)} \left( \eta\lm - \chi\lm - \varphi\lm - \varsigma\lm  \right) Y\lm_{ab} \\
                                          & -\frac{\left(1+z\right)^2}{r a\o} v^{AB} \sum_{(\ell m)}  \nabla_A^{\vphantom{(\ell m)}} k_B\lm \overbar{Y}\lm_{ab}\,.                                          
   \end{split}
   \end{equation}

By construction,  Ricci- and Weyl focussing terms in the optical tidal matrix are expressed as sums over spherical harmonic functions representing its trace $\gamma_{ab} Y\lm$ and trace-free parts ($Y\lm_{ab}$ and $\overbar{Y}\lm_{ab}$). The correction to the Jacobi map can now be decomposed in a similar way. For simplicity, we define an orthonormal set of spherical harmonic basis functions in screen space given by

\begin{align}
 \label{lltb:perturbation:20}
 \gamma_{ab} \tilde{Y}\lm &= \frac{1}{\sqrt{2}} \gamma_{ab} Y\lm \,, \\
 \label{lltb:perturbation:21}
 \tilde{Y}\lm_{ab} &= \sqrt{2\frac{(\ell-2)!}{(\ell+2)!}} Y\lm_{ab}\,, \\
 \label{lltb:perturbation:22}
 \tilde{\overbar{Y}}\lm_{ab} &= \sqrt{2\frac{(\ell-2)!}{(\ell+2)!}} \overbar{Y}\lm_{ab}\,, 
 \end{align}

 that fulfills
 
 \begin{equation}
  \int_{\Omega} \d\Omega \ X\lm_{ab} {Z\lmp}^{ab\ast} = \delta_{XZ} \ \delta_{\ell \ell'} \delta_{mm'}\,,
  \label{lltb:perturbation:23}
 \end{equation}

 with $X$ and $Z$ representing the expressions (\ref{lltb:perturbation:20})-(\ref{lltb:perturbation:22}). \bigskip
 
 The first order correction to the Jacobi matrix can then be written as
 
 \begin{equation}
  D^{(1)}_{ab} = \sum_{(\ell m)} \left( D^{T (\ell m)} \tilde{Y}\lm \gamma_{ab} + D^{TF (\ell m)} \tilde{Y}\lm_{ab} + \overbar{D}\lm \tilde{\overbar{Y}}\lm_{ab} \right) \,.
   \label{lltb:perturbation:24}
 \end{equation}
 
 By projection, we can now obtain the full spherical harmonic decomposition of Eq. (\ref{lltb:perturbation:19}) in this orthonormal harmonic basis:
 
 \begin{align}
  \label{lltb:perturbation:25}
  \begin{split}
  \frac{\d^2 D^{T (\ell m)}}{\d \lambda^2}       &= - \left(1+z\right)^2 \frac{\alpha}{2}\ D^{T (\ell m)} -\left(1+z\right)^2 ra\o \frac{\alpha}{\sqrt{2}} \left( 2w\lm + \Delta\lm + 2\eta\lm \right. \\
                                           &  \left. \quad - \chi\lm - \varphi\lm - \varsigma\lm \right)\,, 
  \end{split}\\
  \label{lltb:perturbation:26}
  \frac{\d^2 D^{TF (\ell m)}}{\d \lambda^2}    &= - \left(1+z\right)^2 \frac{\alpha}{2}\ D^{TF (\ell m)} - \frac{\left(1+z\right)^2}{r a\o} \sqrt{\frac{(\ell+2)!}{2(\ell-2)!}} \left( \eta\lm - \chi\lm - \varphi\lm - \varsigma\lm  \right)\,, \\
  \label{lltb:perturbation:27}
 \frac{\d^2 \overbar{D}\lm}{\d \lambda^2}  &= - \left(1+z\right)^2 \frac{\alpha}{2}\ \overbar{D}\lm -\frac{\left(1+z\right)^2}{r a\o} \sqrt{\frac{(\ell+2)!}{2(\ell-2)!}} v^{AB} \nabla_A^{\vphantom{(\ell m)}} k_B\lm\,.
 \end{align}

 As the full initial shape of the lightcone has to be Minkowskian close to the observer's position, we require vanishing initial conditions at perturbation level:
 
 \begin{equation}
  D^{X(\ell m)}(0) = 0 = \left. \dfrac{\d D^{X(\ell m)}}{\d \lambda}\right|_{\lambda=0} 
 \label{lltb:perturbation:28}
 \end{equation}

 with $X = T, \ TF, \ \overbar{(...)}$.

\section{Green's function to the Jacobi matrix equation and lensing observables}
\label{sect:greens}

Given a generic linear second order initial value problem of the form

\begin{align}
 \label{lltb:greens:1}
 a(t) \ddot{y}(t) + b(t)\dot{y}(t) + c(t)y(t) = f(t)\,, \\
 y(t) = y_0\,, \\
 \dot{y}(t) = v_0\,,
 \end{align}

 it can be shown (see \cite{hermann_partial_2015}) by variation of constants that the Green's function to the linear operator $\mathcal{L} = a(t)\frac{\d^2}{\d t^2}+b(t)\frac{\d}{\d t} +c(t)$ can be expressed in terms of two linearly independent solutions $y_1(t)$ and $y_2(t)$ of the homogeneous Eq. (\ref{lltb:greens:1}). One obtains
 
 \begin{equation}
  y(t) = y_h(t) + \int_0^t{ \d t'\ \frac{y_1(t')y_2(t)-y_1(t)y_2(t')}{a(t') W(t')} \ f(t')}\,,
  \label{lltb:greens:2}
 \end{equation}
 
 where $y_1(0)=0$, $\dot{y}_1(0)\neq0$, $y_2(0)\neq0$, $\dot{y}_2(0)=0$.  \bigskip
 
 This leaves us with the Green's function
 
 \begin{equation}
  G(t,t') =  \frac{y_1(t')y_2(t)-y_1(t)y_2(t')}{a(t') W(t')}
  \label{lltb:greens:3}
 \end{equation}
 
 The Wronskian of the two linearly-independent solutions is given by $W(t)= y_1(t) \dot{y}_2(t) - y_2(t) \dot{y}_1(t)$. \bigskip
 
 In fact, the dynamics of the Wronskian ($\dot{W}(t) = -b(t)\ W(t))$ allows to construct the second linear independent solution $y_2$ from the first one (see \cite{Arfken:379118})
 
 \begin{equation}
   y_2(t) = W(0) y_1(t) \int_0^t \frac{\d t'}{y_1^2(t')} \ \exp{\left(-\int_0^{t'} \d t'' \ b(t'') \right)}\,.    
    \label{lltb:greens:4}
 \end{equation}
 
 The geodesic deviation Eqs. (\ref{lltb:perturbation:25})-(\ref{lltb:perturbation:27}) denote inhomogeneous linear second order ordinary differential equations in the affine parameter $\lambda$. The structure of their homogeneous parts is identical to Eq. (\ref{lltb:background:8}) such that it is solved by 
 
 \begin{equation}
  D_1(\lambda) = r(\lambda) a\o(t(\lambda), r(\lambda)) \equiv (ra\o)(\lambda)\,. 
  \label{lltb:greens:5}
 \end{equation}
 
 Due to the absence of a term $\sim \frac{\d D}{\d \lambda}$ the Wronskian is constant and Eq. (\ref{lltb:greens:4}) simplifies considerably. We then find a possible second, linearly independent solution
 
  \begin{equation}
  D_2(\lambda) = D_1(\lambda) \int_0^\lambda \ \frac{\d \lambda'}{D_1^2(\lambda')} = (ra\o)(\lambda) \int_0^\lambda \frac{\d \lambda'}{(ra\o)^2(\lambda')} \,. 
  \label{lltb:greens:6}
 \end{equation}
 
 According to Eq. (\ref{lltb:greens:3}), the Green's function to the linear operator $\mathcal{L}=\frac{\d^2}{\d\lambda^2}+(1+z(\lambda))^2\frac{\alpha}{2}$ then reads
 
 \begin{equation}
   G(\lambda, \lambda') = (ra\o)(\lambda) (ra\o)(\lambda') \int_{\lambda'}^\lambda \frac{\d \lambda''}{(ra\o)^2(\lambda'')}\,.
   \label{lltb:greens:7}
 \end{equation}
 
Since the initial conditions to the correction to the Jacobi map are trivial (see Eq. (\ref{lltb:perturbation:28})), the homogeneous solution to Eqs. (\ref{lltb:perturbation:25}) - (\ref{lltb:perturbation:27}) is trivial as well. The generic solution is then given by

\begin{equation}
  {D^X}\lm(\lambda) = (ra\o)(\lambda) \int_0^\lambda \d \lambda' (ra\o)(\lambda') \int_{\lambda'}^\lambda \frac{\d \lambda''}{(ra\o)^2(\lambda'')} {F^X}\lm(\lambda')\,,
\label{lltb:greens:8}
\end{equation}

with $X = T, \ TF, \ \overbar{(...)}$ where the latter refers to the axial ``barred" quantity.

The source terms are given by

\begin{align}
 \label{lltb:greens:9}
  F^{T (\ell m)} &= -\left(1+z\right)^2 ra\o \frac{\alpha}{\sqrt{2}} \left( 2w\lm + \Delta\lm + 2\eta\lm - \chi\lm - \varphi\lm - \varsigma\lm \right)\,, \\
 \label{lltb:greens:10}
 F^{TF (\ell m)} &= - \frac{\left(1+z\right)^2}{r a\o} \sqrt{\frac{(\ell+2)!}{2(\ell-2)!}} \left( \eta\lm - \chi\lm - \varphi\lm - \varsigma\lm  \right)\,, \\
 \label{lltb:greens:11}
 \overbar{F}\lm &= -\frac{\left(1+z\right)^2}{r a\o} \sqrt{\frac{(\ell+2)!}{2(\ell-2)!}} v^{AB} \nabla_A^{\vphantom{(\ell m)}} k_B\lm\,.
 \end{align}\bigskip

In the limit of a conformally static FLRW metric

\begin{equation}
 \d s^2 = a^2(\eta) \left(-\d \eta^2 + \d w^2 + f_K(w) \d \Omega^2 \right)\,,
  \label{lltb:greens:12}
\end{equation}

 we can identify w.l.o.g. $\lambda$ with the radial coordinate $w$ using the conformal invariance of null geodesics. Eq. (\ref{lltb:greens:7}) then reduces to 
 
\begin{equation}
 G(w, w') = f_K(w-w')
\label{lltb:greens:13}
\end{equation}

which is the well-known weight function for line-of-sight integrals in weak gravitational lensing (see \cite{bartelmann_topical_2010}). \bigskip

We now apply the definition of the lensing amplification matrix $A_{ab}$ in Eqs. (\ref{lltb:generic:20})-(\ref{lltb:generic:21}) and decompose it into its trace and trace-free parts with respect to the orthonormal harmonic basis defined by Eqs. (\ref{lltb:perturbation:20})-(\ref{lltb:perturbation:22}). This allows to identify the convergence and shear coefficients as

\begin{align}
 \label{lltb:greens:14}
 \kappa\lm(\lambda) &= \frac{D^{T (\ell m)}(\lambda)}{(ra\o)(\lambda)} =  \int_0^\lambda {\d \lambda' \ (ra\o)(\lambda') \int_{\lambda'}^\lambda {\frac{\d \lambda''}{(ra\o)^2(\lambda'')} } F^{T (\ell m)}(\lambda') }\,, \\
 \label{lltb:greens:15}
 \gamma\lm(\lambda) &= \frac{D^{TF (\ell m)}(\lambda)}{(ra\o)(\lambda)} =  \int_0^\lambda \d \lambda' \ (ra\o)(\lambda') \int_{\lambda'}^\lambda \frac{\d \lambda''}{(ra\o)^2(\lambda'')} F^{TF (\ell m)}(\lambda')\,, \\
 \label{lltb:greens:16}
 \overbar{\gamma}\lm(\lambda) &= \frac{\overbar{D}\lm(\lambda)}{(ra\o)(\lambda)} =  \int_0^\lambda \d \lambda' \ (ra\o)(\lambda') \int_{\lambda'}^\lambda \frac{\d \lambda''}{(ra\o)^2(\lambda'')} {\overbar{F}}\lm(\lambda')\,.
 \end{align}

Harmonic powerspectra of lensing observables can then generically be expressed as

\begin{equation}
\label{lltb:greens:17}
\begin{split}
 \langle \tilde{X}\lm(\lambda) \tilde{Z}\lmp(\lambda')^\ast \rangle &= \int_0^\lambda \d x \ (ra\o)(x) \int_0^{\lambda'} \d x' \ (ra\o)(x') \int_x^\lambda \frac{\d y}{(ra\o)^2(y)} \int_{x'}^{\lambda'} \frac{\d y'}{(ra\o)^2(y')} \\
                                               & \quad \langle F^{X(\ell m)}(x) {F^{Z(\ell' m')}(x')}^\ast \rangle \\
                                               &\equiv C^\ell_{\tilde{X}\tilde{Z}} \delta^{\vphantom{\ell}}_{\ell\ell'} \delta^{\vphantom{\ell}}_{mm'} \,, 
\end{split}
\end{equation}

with $\tilde{X}, \tilde{Z} = \kappa, \gamma$ and $X,Z = T, \ TF, \ \overbar{(\ldots)}$. Eq. (\ref{lltb:greens:17}) is a very crucial result as it allows to map the abstract gauge-invariant quantities of linear perturbation theory in $\Lambda$LTB models to actual observable quantities known from weak gravitational lensing. It is therefore conceptually a most welcome tool to constrain $\Lambda$LTB models with information related to linear structure formation.

\section{E- and B-modes for a central observer}
\label{sect:EB-modes}

An alternative harmonic decomposition of Eq. (\ref{lltb:perturbation:19}) that is more commonly applied in weak gravitational lensing as well as CMB studies employs spin-2-weighted spherical harmonics. A generic spin-s spherical harmonic function on the sphere can be defined as

\begin{equation}
 {}_{s} Y\lm = \sqrt{\frac{(\ell-s)!}{(\ell+s)!}} \eth^s Y\lm
 \label{lltb:EB-modes:1}
\end{equation}

using the ``edth" operator $\eth$ (see \cite{goldberg_spin-s_1967, dray_relationship_1985}). \bigskip

By expanding the polar and axial tensor spherical harmonics given in Eqs. (\ref{lltb:perturbation:13}) and (\ref{lltb:perturbation:17}) with respect to the dual helicity basis

\begin{equation*}
 \Theta^{\pm} = \frac{1}{\sqrt{2}} \left(  \d\theta \pm \sin\theta \d \phi \right)\,,
\end{equation*}

we find the correspondence

\begin{equation}
 {}_{\pm 2} Y\lm \left( \Theta^\pm \otimes \Theta^\pm \right)_{ab}  = \sqrt{\frac{(\ell-2)!}{(\ell+2)!}} \left( Y\lm_{ab} \pm \mathrm{i} \overbar{Y}\lm_{ab} \right) \,.
 \label{lltb:EB-modes:2}
\end{equation}

By comparing two spherical harmonic expansions of the shear field using Eq. (\ref{lltb:EB-modes:2}), we can extract the expressions for the E- and B-modes. First of all, we notice that

\begin{align}
 {}_{\pm 2}\gamma\lm = \frac{1}{2} \sqrt{ \frac{(\ell+2)!}{(\ell-2)!}} \left( \gamma\lm \mp \mathrm{i} \overbar{\gamma}\lm\, \right). 
 \label{lltb:EB-modes:3}
\end{align}

Spherical harmonic coefficients of the E- and B-mode signal are rotationally invariant and therefore scalar quantities on $\mathcal{S}^2$ (see for example \cite{bartelmann_topical_2010}). Consequently, we define auxiliary scalar quantities 

\begin{equation}
 \gamma_{\pm}\lm =  \sqrt{\frac{(\ell+2)!}{(\ell-2)!}} {}_{\pm 2} \gamma\lm
\label{lltb:EB-modes:4}
\end{equation}

that arise from applying the edth operator and its complex conjugate twice onto the spin-(-2) and spin-2 shear field, respectively. The spherical harmonic coefficients of the E- and B-mode signal are then given by

\begin{align}
 \label{lltb:EB-modes:5}
 a\lm_{E} &=  \frac{1}{2} \sqrt{\frac{(\ell+2)!}{(\ell-2)!}} \left(\gamma\lm_{+}+\gamma\lm_{-}\right)\,, \\
 \label{lltb:EB-modes:6}
 a\lm_{B} &=  -\frac{\mathrm{i}}{2} \sqrt{\frac{(\ell+2)!}{(\ell-2)!}} \left(\gamma\lm_{+}-\gamma\lm_{-}\right)\,. 
\end{align}

Combining Eqs. (\ref{lltb:EB-modes:3})-(\ref{lltb:EB-modes:6}), we obtain

\begin{align}
 \label{lltb:EB-modes:7}
 a\lm_{E}(\lambda) &=  \frac{1}{2} \sqrt{\frac{(\ell+2)!}{(\ell-2)!}} \gamma\lm(\lambda) =  \frac{1}{2} \sqrt{\frac{(\ell+2)!}{(\ell-2)!}} \int_0^\lambda \d \lambda' \ G(\lambda, \lambda') F^{TF(\ell m)}(\lambda')   \,, \\
 \label{lltb:EB-modes:8}
 a\lm_{B}(\lambda) &=  -\frac{1}{2}\sqrt{\frac{(\ell+2)!}{(\ell-2)!}} \overbar{\gamma}\lm(\lambda) = -\frac{1}{2} \sqrt{\frac{(\ell+2)!}{(\ell-2)!}} \int_0^\lambda \d \lambda' \ G(\lambda, \lambda') \ \overbar{F}\lm(\lambda') \,,
\end{align}

with the Green's function $G(\lambda, \lambda')$ given in Eq. (\ref{lltb:greens:7}). \bigskip

This result is not surprising. The central worldline allows to identify the angular coordinates of the comoving observer with the $\Lambda$LTB angular coordinates. Consequently, the spherical harmonic decomposition of the lensing signal agrees with the one of the gauge-invariant linear perturbations. The E-mode weak lensing signal is therefore exclusively sourced by the polar spherical harmonic branch whereas the B-modes are solely covered by axial perturbations. These results are expected to change if off-center observers are considered since spherical harmonic basis systems then cannot trivially be identified anymore.

\section{Conclusion}
\label{sect:conclusion}

In this paper, we have combined a relativistic formalism of light propagation with gauge-invariant linear perturbation theory in $\Lambda$LTB models. The resulting geodesic deviation (or Sachs) equation allows to map the abstract gauge-invariant quantities describing linear perturbations in $\Lambda$LTB models to actual observables. So far, the analysis is restricted to observers placed at the center of the $\Lambda$LTB patch. Although, conceptually, solutions can be extended to off-center observers, severe technical problems will occur since the initial spherical harmonic expansion of the lensing signal and the $\Lambda$LTB gauge-invariants have to be transformed into each other. We therefore postpone this analysis to a future study. Given a central observer, the geodesic deviation equation can be expanded into the same harmonic basis system as the linear, gauge-invariant perturbations. The resulting system of linear differential equations per spherical harmonic mode $(\ell,m)$ can effectively be solved by a Green's function approach which results in line-of-sight integral expressions analogously to the treatment in FLRW models. Expressions for the convergence and cosmic shear spherical harmonic coefficients have been derived as well as a general expression for their harmonic powerspectra and covariances. In addition, those have been converted into the E- and B-mode contributions to the cosmic shear signal. We found that, due to spherical symmetry of the background solution on the central worldline, axial and polar spherical harmonic modes strictly split into the B- and E-mode contributions, respectively.\bigskip

This work outlines all necessary steps to connect dynamical information from gauge-invariant linear perturbation theory to observable implications on the backward lightcone. It is essential to extend the analysis of $\Lambda$LTB models and especially include constraints from linear structure formation in a consistent manner. By integrating the $\Lambda$LTB master and constraint equations numerically in a cosmologically relevant scenario, we hope to apply this formalism to predict the cosmic shear powerspectrum in realistic $\Lambda$LTB models in the near future. \bigskip

In addition, we hope to, on the one hand, extend the approach to off-center observers and, on the other hand, develop a similar formalism for the integrated Sachs-Wolfe effect in $\Lambda$LTB models. Aiming at a robust test of the Copernican Principle, we hope to put as many constraints as possible onto the density profile of the surrounding universe on Gpc scales.  

\appendix

\section{Appendix: Solution of the Jacobi matrix equation on the background level}
\label{app:solution}

This section contains a short proof that the areal radius indeed solves the Jacobi matrix equation at the background level. Interestingly, this result can be obtained without exact knowledge of the shape of the backward lightcone since only differential relations between lightcone coordinates, redshift and affine parameter are going to enter. \bigskip

We start again from the background $\Lambda$LTB metric,

\begin{equation}
 \d s^2 = -\d t^2 + \frac{a\p^2(t,r)}{1-\kappa(r) r^2} \d r^2 + r^2 a\o^2(t,r) \d \Omega^2\,,
 \label{lltb:app:solution:1}
\end{equation}

and the energy-momentum tensor $T_{\mu\nu} = \rho(t,r) u_\mu u_\nu$. Introducing the free function $M(r)$ (see for example \cite{plebanski_introduction_2012}), Einstein's field equations can be reduced to two remaining expressions 

\begin{align}
   \label{lltb:app:solution:2}
   \frac{(r^3 M(r))'}{r^2 a\o^2 a\p} = 8\pi G \rho\,, \\
   \label{lltb:app:solution:3}
   H\o^2 = \frac{M(r)}{a\o^3} - \frac{\kappa(r)}{a\o^2} + \frac{\Lambda}{3}\,.
\end{align}

Following (\cite{clarkson_perturbation_2009}), we define an auxiliary function $W(t,r)$ and a so-called radial frame derivative given by

\begin{align}
 \label{lltb:app:solution:4}
 W(t,r) &:= \frac{\sqrt{1-\kappa(r) r^2}}{r a\o(t,r)}\,, \\
 \label{lltb:app:solution:5}
 (\ldots)' &:= \frac{\sqrt{1-\kappa(r)r^2}}{a\p(t,r)} \partial_r (\ldots)\,.  
\end{align}

Within this notation, Eq. (\ref{lltb:app:solution:2}) can be transformed into an equivalent expression 

\begin{equation}
 W' = -W^2 - 4\pi G \rho + H\o H\p + \frac{M}{2 a\o^3} - \frac{\Lambda}{3} 
\label{lltb:app:solution:6}
\end{equation}

involving the tangential and radial Hubble rates $H\o=\dot{a\o}/a\o$ and $H\p=\dot{a\p}/a\p$. \bigskip

We now reconsider the Jacobi matrix equation for central observers in $\Lambda$LTB spacetimes

\begin{equation}
 \frac{\d^2 D_{ab}(\lambda)}{\d \lambda^2} = - 4\pi G \rho\left[t(\lambda), r(\lambda)\right] \left[1+z(\lambda)\right]^2 D_{ab}(\lambda)\,.
 \label{lltb:app:solution:7}
\end{equation}

Inserting $D_{ab} = r a\o(t,r) \gamma_{ab}$ into Eq. (\ref{lltb:app:solution:7}), we find
  
  \begin{equation}
   \begin{split}
   \frac{\d^2}{\d \lambda^2} \left\{r(\lambda) a\o\left[t(\lambda), r(\lambda)\right]\right\} = & \ (1+z)^2 H\p (\sqrt{1-\kappa r^2} - r \dot{a}\o) \\
                                                                        + & \ (1+z)^2 ra\o \left[\frac{\sqrt{1 - \kappa r^2}}{a\p} \partial_r \left(\frac{\sqrt{1-\kappa r^2}}{r a\o}\right) + \frac{1-\kappa r^2}{r^2 a\o^2}\right] \\
                                                                        - & \ (1+z)^2 \sqrt{1-\kappa r^2} H\p + (1+z)^2 r \ddot{a}\o \\
                                                          \overset{!}{=} & \ 4\pi G \rho (1+z)^2 r a\o \,,
 \label{lltb:app:solution:8}
  \end{split}
  \end{equation}

  where the differential relations for redshift and lightcone coordinates with respect to $\lambda$ have been applied (see Eqs. (\ref{lltb:background:4}) - (\ref{lltb:background:6})).\bigskip

  Using Eqs. (\ref{lltb:app:solution:2}) and (\ref{lltb:app:solution:3}) as well as the definitions of the radial and tangential Hubble rates, the Jacobi matrix equation can, after some algebra, be transformed into
  
  \begin{equation}
    W' + W^2 - H\p H\o + \frac{\ddot{a}\o}{a\o} = -4\pi G \rho\,, 
 \label{lltb:app:solution:9}
  \end{equation}

   which is very close to Eq. (\ref{lltb:app:solution:5}). In fact, differentiating Eq. (\ref{lltb:app:solution:3}) with respect to time yields

 \begin{equation}
  \frac{\ddot{a}\o}{a\o} = - \frac{M}{2a\o^3} + \frac{\Lambda}{3}\,.
 \label{lltb:app:solution:10}
  \end{equation}

  Eq. (\ref{lltb:app:solution:9}) can now be replaced and we finally obtain Eq. (\ref{lltb:app:solution:6}). Thus, the Jacobi matrix equation has been transformed to a well-known relation from Einstein's field equations, once the areal radius is inserted. In order to uniquely identify $D_{ab}(\lambda)$ with $r(\lambda) a\o\left[t(\lambda), r(\lambda)\right] \gamma_{ab}$, the initial conditions need to coincide as well. Since $r(\lambda=0)=0$ and $a\o(t,r)$ only weakly depends on $r$ close to the center of the $\Lambda$LTB patch \footnote{As alternative, physical explanation it can be mentioned that a central, freely-falling observer locally experiences a Minkowski spacetime which automatically implies $\left. r a\o(t,r) \right|_{\lambda=0} = 0 = r(\lambda=0)$.}, we have 
   
   \begin{align}
    \label{lltb:app:solution:12}
    \left. r a\o(t,r) \right|_{\lambda=0} &= 0\,, \\
     \label{lltb:app:solution:13}
    \begin{split}
    \left. \frac{\d}{\d \lambda} r a\o(t,r) \right|_{\lambda=0} &= \partial_r\left[r a\o(t,r)\right] \left. \frac{\d r}{\d \lambda}\right|_{\lambda=0} + \partial_t \left[r a\o(t,r)\right] \left. \frac{\d t}{\d \lambda}\right|_{\lambda=0}  \\
                                                                &= \sqrt{1-\kappa(0) r^2(0)} \left[1+z(0)\right] - r(0) a\o(t_\mathrm{age}, 0)  \left[1+z(0)\right] \\
                                                                &= 1\,. 
    \end{split}
  \end{align}

  Hence, $D_{ab}(\lambda) = r(\lambda) a\o\left[t(\lambda), r(\lambda)\right] \gamma_{ab}$ uniquely solves the Jacobi matrix equation for central, freely-falling observers in generic $\Lambda$LTB spacetimes.

 \acknowledgments
 We thank  Matthias Redlich, Simon Hirscher and Bj\"{o}rn-Malte Sch\"{a}fer for extensive discussions, support and encouragement especially at the beginning of this work. This project has been supported by the German \emph{Deut\-sche For\-schungs\-ge\-mein\-schaft, DFG\/} project number BA 1369 / 20-2. 
 

\bibliographystyle{JHEP}
\bibliography{references/refs}

\end{document}